\documentstyle[epsfig]{mn}

\def\gs{\mathrel{\raise0.35ex\hbox{$\scriptstyle >$}\kern-0.6em
\lower0.40ex\hbox{{$\scriptstyle \sim$}}}}
\def\ls{\mathrel{\raise0.35ex\hbox{$\scriptstyle <$}\kern-0.6em
\lower0.40ex\hbox{{$\scriptstyle \sim$}}}}

\begin{document}

\title[GRBs and star formation] 
{Gamma-ray bursts and the history 
of star formation} 

\author 
[A.\,W. Blain and Priyamvada Natarajan] 
{ 
A. W. Blain$^{1,2}$
and Priyamvada Natarajan$^{1,3}$\\
\vspace*{1mm}\\
$^1$ Institute of Astronomy, Madingley Road, Cambridge, CB3 0HA\\
$^2$ Cavendish Laboratory, Madingley Road, Cambridge, CB3 0HE\\ 
$^3$ Department of Astronomy, Yale University, New Haven, CT 06520-8181, USA\\ 
} 
\maketitle 

\begin{abstract} 
Popular models for the origin of gamma-ray bursts (GRBs) include
short-lived massive stars as the progenitors of the fireballs.  Hence
the redshift distribution of GRBs should track the cosmic
star formation rate of massive stars accurately. A significant
proportion of high-mass star formation activity appears to occur in
regions that are obscured from view in the optical waveband by
interstellar dust.  The amount of dust-enshrouded star formation
activity taking place has been estimated by observing the thermal
radiation from the dust that has been heated by young stars in the
far-infrared and submillimetre wavebands. Here we discuss an
alternative probe -- the redshift distribution of GRBs. GRBs are
detectable at the highest redshifts, and because gamma rays are not
absorbed by dust, the redshift distribution of GRBs should therefore
be unaffected by dust extinction. At present the redshifts of GRBs can
only be determined from the associated optical transient emission;
however, useful information about the prevalence of dust-obscured star
formation can also be obtained from the ratio of GRBs with and without
an associated optical transient. Eight GRBs currently have
spectroscopic redshifts. Once about a hundred redshifts are known, the
population of GRBs will provide an important test of different models
of the star formation history of the Universe.
\end{abstract} 


\section{Introduction} 

Gamma-ray bursts (GRBs) are detectable out to the edges of the
observable Universe, and so provide information about the processes
occurring within their progenitors at all cosmic epochs (Piran 1999a,b).
If GRBs arise either from binary mergers of massive stellar remnants
(neutron stars and black holes; Paczynski 1986) or from failed
supernovae collapsing to form black holes (Woosley 1993), or from the
collapse of massive stellar cores (hypernovae; Paczynski 1998), then
they will be associated with the formation of massive stars. See 
Hogg \& Fruchter (1999) and Holland \& Hjorth (1999) for some 
observational evidence that this is the case. Because
high-mass stars have very short lifetimes, the rate of GRBs should
trace the formation rate of massive stars in the Universe. Hence, if
the distribution of the redshifts of GRBs is known, this should allow
the evolution of the high-mass star-formation rate to be derived
(Totani et al.\ 1997; Wijers et al.\ 1998; Hogg \& Fruchter 1999; 
Mao \& Mo 1999; Krumhotz, 
Thorsett \& Harrison 1999).

Observations of faint galaxies in the optical and near-infrared
wavebands have been used to estimate the history of star formation
activity (Lilly et al.\ 1996; Madau et al.\ 1996; Steidel et al.\
1999); however, absorption by interstellar dust in star-forming
galaxies could significantly modify the results (Blain et al.\
1999a,c). It is difficult but not impossible to correct these
optically derived results to take account of this effect. By making
observations in the near-infrared waveband, where the optical depth of
the dust is less, some progress has been made (Pettini et al.\ 1998;
Steidel et al.\ 1999). However, there are considerable uncertainties
in the size of the corrections that should be applied.  It is now
possible to detect the energy that has been absorbed and re-emitted by
dust in high-redshift galaxies directly in the form of thermal
far-infrared radiation, which is redshifted into the submillimetre
waveband. The sensitive SCUBA camera at the James Clerk Maxwell
Telescope has revealed this emission directly (Smail, Ivison \& Blain
1997; Barger et al.\ 1998; 
Hughes et al.\ 1998; Blain et al.\ 1999b, 2000; 
Barger, Cowie \& Sanders 1999; Eales et al.\ 1999). It is possible to derive a
history of high-mass star formation from the SCUBA observations (Blain
et al.\ 1999a,c); at least as much energy is inferred to have been
released from dust-enshrouded star formation activity as in the form
of unobscured starlight. An independent test of the relative amount of
obscured and unobscured star formation activity would be extremely
valuable.

The advantage of using GRBs for this purpose is that gamma-rays are
not absorbed by interstellar dust, either within the host galaxy of
the GRB or in the intergalactic medium along the line of sight to the
host galaxy, and so dust extinction will not modify our view of the
distant Universe observed using gamma rays. However, there is a
problem, as in order to determine a redshift for a GRB the spectrum of
the associated burst of optical transient radiation must be
detected. If the burst is heavily enshrouded in dust, then it would be
difficult to detect such a transient and to obtain its spectrum.

In Section\,2 we briefly review the differences between the histories of 
star formation derived from optical/near-infrared and 
far-infrared/submillimetre observations. In Section\,3 we discuss and 
predict the associated redshift distribution of GRBs. In Section\,4 we
discuss the consequences for determining the history of star formation 
using observations of GRBs. We 
assume an Einstein--de Sitter world model with Hubble's constant 
$H_0=50$\,km\,s$^{-1}$\,Mpc$^{-1}$.

\section{The history of star formation} 

In Fig.\,1 we compare five currently plausible star formation
histories (Blain et al.\ 1999a,c), two of which are based on
observations made in the optical and near-infrared wavebands and three
of which are based on observations in the far-infrared and
submillimetre wavebands. A wide variety of observational data that has
been gathered in the optical and near-infrared wavebands is also
plotted. The data is described in more detail in the caption of
fig.\,1 of Blain et al.\ (1999a) and by Steidel et al.\ (1999).

In the first optically derived model (thin dashed line) it is assumed
that no dust absorption takes place within the galaxies detected in
deep optical images, and that there is no population of strongly
obscured objects missing from these samples. This model closely
follows the form of the history of star formation derived by Madau et
al.\ (1996) from an analysis of the {\it Hubble Deep Field} at $z \gs
2$, and from observations of the Canada--France Redshift Survey fields
at $z \ls 1 $ by Lilly et al. (1996). In the second optically derived
model (thick dashed line), dust extinction is assumed to be
present. The model fits the data that has been corrected empirically
to take account of the effects of dust, using radio and {\it
ISO} satellite observations at $z \ls 1$ (Flores et al.\ 1999) and
using estimates of extinction in high-redshift galaxies estimated from
near-infrared spectroscopy (Pettini et al.\ 1998; Steidel et al.\
1999).

The three different far-infrared/submillimetre models of the history of star 
formation fit all of the available data describing the background radiation 
intensity and the counts of dusty galaxies in these wavebands. The 
`Gaussian model' (thin solid line) was derived by Blain et al.\ (1999c), and 
the `Modified Gaussian model' (medium solid line) was changed slightly 
in order to fit the median redshift of plausible counterparts to 
submillimetre-selected galaxies (Smail et al.\ 1998) determined by 
Barger et al.\ (1999b). In the `Hierarchical model' (thick solid line)  
the population of submillimetre-luminous galaxies is described in 
terms of short-lived luminous bursts induced by galaxy 
mergers (Blain et al.\ 1999a). This model provides a reasonable fit to the 
Barger et al.\ (1999b) redshift distribution if the dust temperature is assumed 
to be 35\,K.

\begin{figure}
\begin{center}
\epsfig{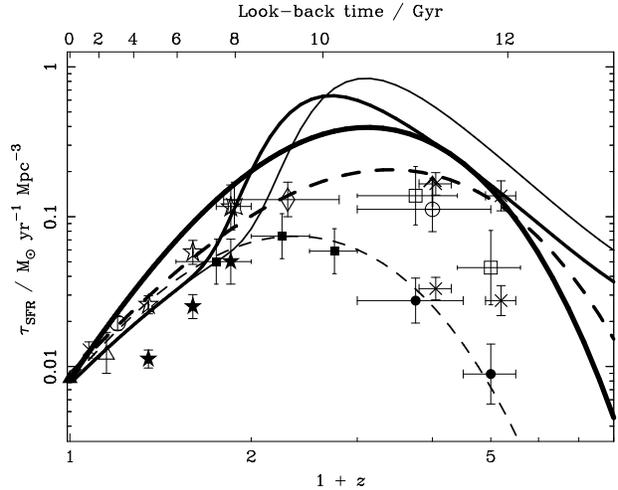}
\end{center}
\caption{ 
A summary of the current state of knowledge of the star formation 
history of the Universe. The data points plotted are described in
Blain et al.\ (1999a) and Steidel et al.\ (1999). The thick 
and thin dashed curves describe models that represent 
the dust-corrected and non-dust-corrected 
histories derived from optical and near-infrared observations, which are 
represented by data points. Where corrections have been made to the 
optical data to account for the estimated effects of dust extinction, the 
data is represented by empty symbols, and the higher pair of 
high-redshift diagonal crosses. Where no corrections have been applied, 
the data is represented by filled symbols and by the lower pair of 
high-redshift diagonal crosses. The solid lines represent 
models derived from far-infrared and submillimetre-wave 
emission: a `Gaussian model' (Blain et al.\ 1999c), a `modified Gaussian 
model' (Barger et al.\ 1999b) and an `hierarchical model' (Blain et al.\ 
1999a), in order of increasing thickness.
}
\end{figure}

\section{The redshift distribution of GRBs}

A histogram showing the eight spectroscopic redshifts of optical 
transients associated with GRBs (Greiner 1999) is plotted in Fig.\,2 
(solid histogram; Metzger et al.\ 1997; Djorgovski et al.\ 1998a,b,c, 
1999a,b; Kulkarni et al.\ 1998; Galama et al.\ 1999; Vreeswijk et al.\ 
1999). There are indications of the redshifts for three more GRBs, 
which are included in the derivation of the dotted histogram shown 
in Fig.\,2, while for about another twenty GRBs no optical transient has been 
detected, in some cases despite sensitive searches. These numbers were 
compiled on 1999 November 16. 

In Fig.\,2 we also present the expected redshift distributions of 
GRBs for each of the five star formation histories shown in Fig.\,1.  
These redshift distributions are derived by integrating the function 
that describes the evolution of the global star-formation rate along 
a radial section of the Universe (Totani 1997; Wijers et al.\ 1998) and 
have been normalised to unity.  

In this calculation we assume that a typical GRB would be detectable at 
any redshift. If, in fact, there is a redshift-dependent selection function 
for GRBs, then the observed redshift distribution of bursts would be expected 
to be systematically lower as compared with the curves shown in Fig.\,2. 
At present, there are too few redshifts with which to estimate the 
possible size of this effect; however, the detection of two GRBs at $z > 3$ 
tends to argue against the strong anti-selection of high-redshift bursts. 

\begin{figure}
\begin{center}
\epsfig{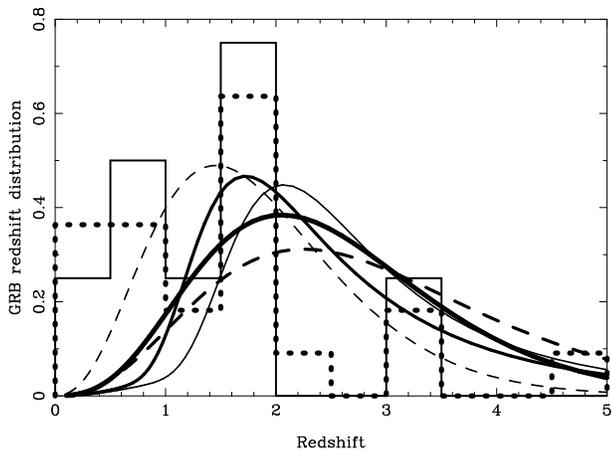}
\end{center}
\caption 
{The curves represent the GRB redshift distributions expected in 
the five star formation histories shown by the curves in Fig.\,1 (the line 
styles are the same as those used in Fig.\,1). The solid histogram shows the 
redshift distribution of the eight GRBs for which optical transients have 
confirmed spectroscopic redshifts (Greiner 1999). It is plausible that a
further three GRBs, GRB970228, GRB970828 and GRB980329
have redshifts $1.3 < z < 2.5$, $z \simeq 0.33$ and $z \sim 5$
respectively (Djorgovski et al.\ 1998b; Greiner 1999; Fruchter 1999). 
GRB980329 is the only GRB that has currently been detected at a 
submillimetre wavelength of 850\,$\mu$m (Smith et al.\ 1999). 
These three sources are included in the dotted
histogram. GRB980424, which has been identified with a low-redshift
supernova, has not been included in the figure.}
\end{figure}

\section{Discussion} 

It is clear from the histograms shown in Fig.\,2 that too few 
redshifts are currently known to allow us to discriminate between the
different models of the star formation history. Nor is the efficiency 
of generating GRBs from massive star formation activity 
known in sufficient detail to allow us to discriminate between the 
different models on the grounds of the absolute number of bursts 
observed. As an additional caveat, it seems likely that about 
20\,per cent of the submillimetre-selected galaxies are powered by 
gravitational accretion onto active galactic nuclei (AGNs; Almaini, 
Lawrence \& Boyle 1999), and as such should not be associated with 
GRBs derived from exploding high-mass stars, unless a significant 
amount of high-mass star formation activity is taking place coevally 
with AGN fueling.

If much of the
high-mass star formation activity in the Universe does indeed take 
place in dust-enshrouded galaxies, then the optical transients of GRBs 
that occur in these galaxies would be less likely to be detected than 
those in dust-free galaxies, because while the GRB gamma-ray signal 
can escape, the associated optical transient emission will be
obscured. Therefore, the inferred GRB redshift distribution at present
might in fact be biased to lower redshifts, given the large number of GRBs 
that do not have detected optical transient counterparts. This selection 
effect for the detection of optical transients will depend on the 
geometry and environment of the host galaxy, and so is likely to be difficult to
investigate reliably until a much larger sample of high-quality follow-up 
observations of GRBs has been assembled.

It is interesting to note that the predicted redshift distribution of GRBs 
that was derived from unobscured optical observations of the star formation 
history, as shown by the thin dashed curve in Fig.\,2, appears to provide 
the best agreement with the observed redshift distribution of the optical 
transients that has been determined so far. This suggests that there could 
be a common extinction bias against the detection of both dust-enshrouded 
star formation activity in optical galaxy surveys and of the optical 
transients of GRBs.

Since systematic, rapid, deep and reliable searches for optical
transients began in 1998 March (Akerlof et al.\ 1999), 
nine gamma-ray bursts have been detected with optical transients, four 
without and a further thirteen unreported, on 1999 November 16 
(Greiner 1999). If those GRBs without optical
transients are in dusty regions, then this lends
support to the idea that comparable amounts of high-mass star
formation occurs in heavily and lightly dust-enshrouded regions. 
One GRB (GRB980329) has so far been detected using SCUBA 
at a submillimetre wavelength of 850\,$\mu$m (Smith et al.\ 1999), 
although it is unclear 
whether there is any component of thermal dust emission involved, 
or if the emission is entirely attributable to synchrotron radiation 
from the shocked interstellar medium. See Taylor et al.\ (1998, 1999) 
for a discussion of the properties of GRBs in the radio waveband, where 
they are not subject to obscuration by dust.  

The number of spectroscopic redshifts for the optical transients of 
GRBs is growing steadily, with a new determination being reported 
every two to three months (Greiner 1999). The number of redshifts for 
GRBs already exceeds the number of redshifts that have been 
obtained for galaxies detected in submillimetre-wave surveys that have 
reliable identifications. Once a sample of about 100 
GRBs have redshifts then it should be possible to discriminate between the
different models. As the predicted redshift distributions shown in Fig.\,2 
are significantly different, the GRB redshift distribution may provide 
a very significant constraint to the history of star formation
activity at all epochs. 

\subsection{Dust-enshrouded infrared transients} 

It will be important to pay close attention to the GRBs without optical 
transients. If the optical and ultraviolet radiation released by the GRB is 
obscured by dust, then since these heated dust grains have a low heat capacity, 
the reprocessed thermal emission could be detected directly as a transient signal 
at submillimetre to near-infrared wavelengths. Intense shocking and heating of the 
gas in the interstellar medium of the host galaxy could also lead to detectable 
emission of far-infrared and submillimetre-wave fine-structure atomic line 
radiation, and molecular rotational line radiation (see also Perna, Raymond \& 
Loeb 2000). Because the host galaxy of the GRB is likely to be optically 
thin at far-infrared and submillimetre wavelengths, the detection of this 
radiation could provide a redshift for a GRB, even in the absence of an optical 
transient source if the opacity at optical wavelengths is very high. In the 
future, the large collecting area, excellent sensitivity and subarcsecond angular 
resolution of the Atacama Large Millimeter Array (ALMA; Wootten 2000) will 
make it the ideal instrument to conduct observations to search for any 
transient line and continuum radiation from GRBs in the submillimetre waveband. 
Recently, Waxman \& Draine (2000) discussed the effects of a GRB on the 
sublimation of dust in the surrounding interstellar medium. We are currently 
investigating the observability of infrared transients from dust-enshrouded 
GRBs (Venemans \& Blain, in preparation).

\section{Conclusions} 

A large statistical sample of redshifts for gamma-ray bursts (GRBs)
will allow the star formation history of the Universe to be probed in
more detail than is currently possible. Once about a hundred examples
are known, the fraction of dust-enshrouded star formation activity that
takes place as a function of redshift can be addressed by an analysis
of both the redshift distribution of the optical transients and the
fraction of GRBs for which an optical transient is detected. The
results will also allow the fraction of AGN in submillimetre-selected
samples and the form of their evolution to be estimated in a new
way. The commissioning of the ALMA interferometer array will hopefully
allow the absorbed optical and ultraviolet light from dust-enshrouded
GRBs to be detected in the form of a far-infrared transient signal,
potentially revealing the redshift of the GRB without requiring
observations in the optical waveband.

\section*{Acknowledgements}

AWB, Raymond \& Beverly Sackler Foundation Research Fellow, gratefully 
acknowledges support from the Foundation as part of the Deep Sky 
Initiative programme at the IoA. We thank Brad Hansen, 
Kate Quirk and an anonymous 
referee for helpful comments on the manuscript, and Andrew Fruchter, 
George Djorgovski and Shri Kulkarni for useful conversations.

\end{document}